\documentclass{he_symp}
\usepackage{psfig,graphicx,epsfig}
\usepackage{color}
\usepackage{amsmath,amssymb,epic,eepic,array}
\begin{document}
\renewcommand{\FirstPageOfPaper }{ 54}\renewcommand{\LastPageOfPaper }{ 57}

\title{PSR B1951+32 and PSR J0537-6910 - where are the optical counterparts?}
\author{R. F. Butler\inst{1}, A. Golden\inst{1}, A. Shearer\inst{1} \and C. Gouiffes\inst{2}}  
\institute{National University of Ireland, Galway; University Road, Galway, Ireland
\and  CEA/DSM/DAPNIA/Service d'Astrophysique, Saclay, Paris, France}

\authorrunning{R. F. Butler et al.}
\maketitle

\begin{abstract}

There remain several definitive $\gamma$-ray pulsars that are as yet
undetected in the optical regime. A classic case is the pulsar PSR
B1951+32, associated with the complex CTB 80 SNR. Previous ground
based high speed 2-d optical studies have ruled out candidates to
$m_{V}$ $\sim$ 24.  Hester (2000) has reported an analysis of archival
HST/WFPC2 observations of the CTB 80 complex which suggest a compact
synchrotron nebula coincident with the pulsar's radio position.
Performing a similar analysis, we have identified a possible optical
counterpart within this synchrotron nebula at $m_{V}$ $\sim$ 25.5, and
another optical counterpart candidate nearby at $m_{V}$ $\sim$24.2.  

PSR J0537-6910 is a young (canonical age $\approx$ 5000 years), 16ms
pulsar in the LMC. We report a search for optical pulsations from the
region around the X-ray position. We see no obvious candidate
exhibiting optical pulsations at the X-ray period, with a 3$\sigma$
upper limit of $m_V \approx $23.6. We have also examined recent
Chandra results (Wang et al. 2001) and show that their X-ray-Optical
astrometry is in error by about 7''.

\end{abstract}

\section{Introduction}

The detection of non-thermal high energy magnetospheric emission from
isolated pulsars has remained a non-trivial problem, despite great
advances in instrumentation and technological expertise.  To date,
only 7 optical pulsars have been detected with emission believed to be
magnetopsherically dominated, and despite considerable effort, only 8
$\gamma$-ray pulsars.  In contrast to radio emission, which is
generally believed to be generated in close proximity to the magnetic
poles, no clear theoretical model construct exists as regards the
higher energies. The two principal schools of thought place
$\gamma$-ray emission localized either to the magnetic poles
(Daugherty \& Harding 1996) or located further out in the
magnetosphere (Romani 1996). Considerable problems remain with these
two frameworks, in terms of predicted fluxes, spectral indices and
light curve morphologies, and it is clear that further work is
required on this subject.  This is all the more relevant when one
attempts to address the growing empirical database of lower energy
emission, in particular in the optical regime. A consequence of
non-linear processes within the magnetosphere, this synchrotron
emission forms a useful constraint with which one can attempt to
comprehensively develop a self-consistent theoretical framework.

\section{PSR B1951+32}

The pulsar PSR B1951+32, located within the complex combination
supernova remnant (SNR) CTB 80, was first identified as a
steep-spectrum, point-like source in the radio (Strom 1987), and
discovery of radio pulses with an unusually fast 39.5-s period quickly
followed (Kulkarni et al. 1988).  Canonically, the pulsar's age and
the estimated dynamical age of the SNR are consistent at $\sim$10$^5$
yrs (Koo et al. 1990) and both have been determined to be at a
distance of $\sim$ 2.5 kpc.  There is thus general agreement that the
association is valid.  Evidence for pulsed emission was subsequently
found in gamma-rays (Ramanamurthy et al. 1995) and possibly in X-rays
(Safi-Harb et al. 1995; Chang \& Ho 1997). The ROSAT observations
in the X-ray regime do indicate a complex light curve strongly
dominated by the intense X-ray radiation of a pulsar-powered
synchrotron nebula (Safi-Harb et al. 1995; Becker \& Truemper
1996). The unambiguous double-peaked $\gamma$-ray (EGRET) light curve
obtained by Ramanamurthy et al.  (1995) at the appropriate spin-down
ephemeris suggested that the pulsar had a conversion efficiency, in
terms of rotational energy to $\gamma$-rays, of $\sim$
0.004. Consequently there are strong grounds for the possibility of an
optical detection. Using a ground-based MAMA detector in the TRIFFID
camera, we have previously examined the central field of CTB 80, but
could find no evidence of pulsations in either B or V from this
region (O'Sullivan et al., 1998). 

\subsection{Analysis of archival HST/WFPC2 observations}

We obtained from the HST archive all existing WFPC2 data of the CTB 80
SNR, as listed in Table~\ref{hst_archive}. The core of the CTB 80
remnant lies on chip WF3 of the WFPC2 camera in every case. Image
processing and photometry were performed using the IRAF (Tody 1993),
STSDAS, and DAOPHOT-II (Stetson 1994) packages. The images in each
band were stacked and cleaned of cosmic rays and hot pixels using
standard techniques. The F547M intermediate-width band enabled us to
perform a deep photometric search for faint stellar sources, to S/N=2
at MAG$_{F547M}$ = 26.7. More details of our reduction procedure,
including refinement of the HST astrometry using the 2MASS
point-source catalog, can be found in Butler et al. (2002). The
photometric and astrometric results are shown in Figure 1 and Table 2.

\begin{table}
      \caption{List of WFPC2 observations of the CTB 80 SNR, obtained from the ST-ECF HST archive.}
         \label{hst_archive}
      \[
         \begin{array}{cccc}
            \hline
            \noalign{\smallskip}
Date & Filter & Total Exptime  & Notes \\
            \noalign{\smallskip}
            \hline
            \noalign{\smallskip}
2/10/97 & F656N & 5300 & H II \\
2/10/97 & F673N & 5400 & S II \\
2/10/97 & F502N & 5400 & O III \\
2/10/97 & F547M & 2600 & \mbox{Str\"omgren} $y$ \\
            \noalign{\smallskip}
             \hline
         \end{array}
      \]
   \end{table} 
  
\begin{table}
      \caption{The positions and magnitudes derived by DAOPHOT-II/allstar
PSF-fitting for point sources detected in the WFPC2 F547M image of CTB
80. The ``Dist.'' column contains the distance from from each source to
the centre of the synchrotron nebula; only sources within 5.0''
of this position are listed here.}
         \label{hst_mags}
      \[
         \begin{array}{ccccc}
            \hline
            \noalign{\smallskip}
 \# & RA (2000) & Dec (2000) & Dist. & MAG_{F547M} \\
  & hh:mm:ss.ss & dd:mm:ss.s & arcsec & magnitudes\\
            \noalign{\smallskip}
            \hline
            \noalign{\smallskip}
              1_{HST} & 19:52:58.24 &  32:52:41.0 &         0.20 &        23.93 \pm 0.30 \\
              2_{HST} & 19:52:58.25 &  32:52:41.8 &         0.97 &        22.15 \pm 0.07 \\
              3_{HST} & 19:52:58.23 &  32:52:42.0 &         1.11 &        21.35 \pm 0.07 \\
              4_{HST} & 19:52:58.28 &  32:52:39.9 &         1.16 &        24.21 \pm 0.12 \\
              5_{HST} & 19:52:58.14 &  32:52:39.0 &         2.23 &        24.07 \pm 0.12 \\
              6_{HST} & 19:52:58.14 &  32:52:43.4 &         2.69 &        21.84 \pm 0.07 \\
              7_{HST} & 19:52:58.42 &  32:52:39.2 &         2.90 &        21.61 \pm 0.06 \\
              8_{HST} & 19:52:58.39 &  32:52:38.7 &         2.95 &        21.96 \pm 0.07 \\
              9_{HST} & 19:52:58.48 &  32:52:42.6 &         3.55 &        23.00 \pm 0.09 \\
             10_{HST} & 19:52:57.95 &  32:52:40.8 &         3.57 &        25.73 \pm 0.26 \\
             11_{HST} & 19:52:58.39 &  32:52:44.1 &         3.73 &        24.25 \pm 0.14 \\
             12_{HST} & 19:52:57.97 &  32:52:39.1 &         3.79 &        25.87 \pm 0.47 \\
             13_{HST} & 19:52:58.02 &  32:52:37.1 &         4.68 &        22.90 \pm 0.07 \\
             14_{HST} & 19:52:58.38 &  32:52:36.6 &         4.74 &        24.44 \pm 0.18 \\
            \noalign{\smallskip}
             \hline
         \end{array}
      \]
   \end{table} 
  
\begin{figure}
\centerline{\psfig{file=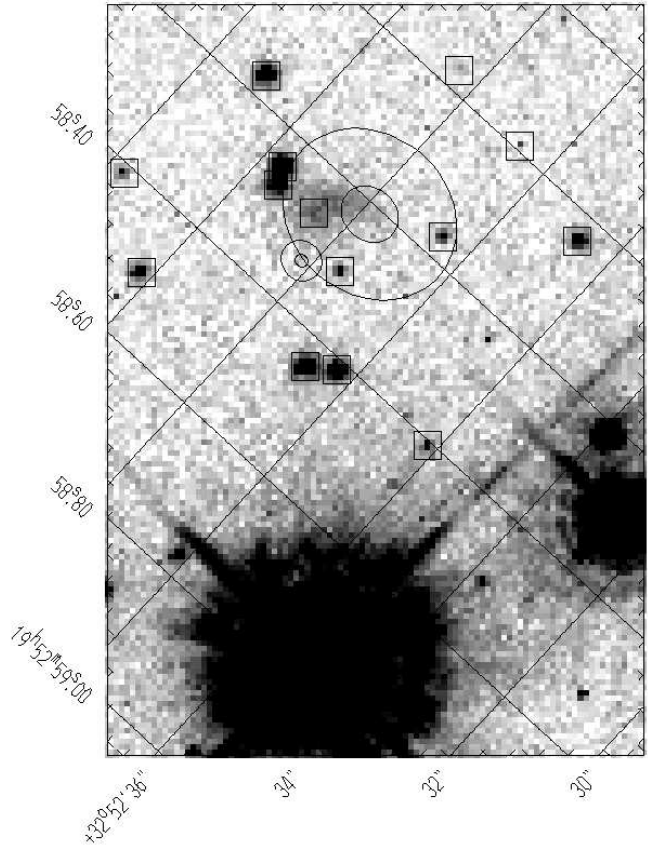,width=8.8cm,clip=} }
\caption{Section of the HST/WFPC2 F547M image of the core of
CTB80. The compact synchrotron nebular knot reported by Hester (2000)
can be seen at the upper-centre.  The coordinate grid shows our
improved astrometric calibration after referencing to the 2MASS
point-source catalog. The mapped radio positions for PSR B1951+32 (see
text) are marked by black ellipses, the semi-major and semi-minor axes
of which are determined by the 1-$\sigma$ and 3-$\sigma$ total
positional uncertainties in RA and Dec - ie. the quoted error on the
radio position, combined with the total rms error on the HST-2MASS fit
for 31 stars. The larger pair of ellipses mark the interferometric
position. The positions derived by DAOPHOT-II/allstar PSF-fitting for
all measured point sources within 5.0'' of the centre of the
synchrotron nebula are indicated by black squares: counterpart
candidate 1$_{HST}$ lies within the nebula and candidate 4$_{HST}$ lies
just below it.
\label{image}}
\end{figure}

The two ``best'' radio positions for PSR B1951+32, from Foster et
al. (1990) [interferometric] and Foster et al. (1994) [glitchless
timing], are shown overplotting the HST field and its detected
point-source content in Figure 1. Two point-sources were found to be
the most positionally consistent with these two radio positions; the
first (object 4$_{HST}$ in Table 2) is a straightforward point-source
measurement at MAG$_{F547M}$ = 24.21 $\pm$ 0.12; whereas the second
(object 1$_{HST}$) is superimposed on the compact synchrotron knot
reported by Hester (2000), and required photometric simulations to
determine its true magnitude, at MAG$_{F547M}$ between 25.0 -
26.0. Further Monte Carlo simulations to determine the limiting magnitude of the
WFPC2 observation showed that an object within this magnitude range is
indeed detectable against such nebular background in this data, albeit
with $S/N$ of only $\approx$2 - 2.5.

Crucially, the corrected magnitude of MAG$_{F547M}$ = 25.0 - 26.0 for
object 1$_{HST}$ is within the range predicted by the successful mode framework
of Pacini \& Salvati (1987) and more recently the phenomenological
analysis of Shearer \& Golden (2001).  The magnitude of object
4$_{HST}$ is also consistent with these models, at the brighter end of
the predicted range. Consequently, taken together with their positions
with respect to the two radio-position error ellipses, we suggest that
these two objects are plausible new optical counterpart candidates to
PSR B1951+32.

\section{PSR J0537-6910}

PSR J0537-6910 was first observed by RXTE satellite (Marshall et
al. 1998). ROSAT observations by Wang et al. (1998) showed that the
pulsar lies within the SNR N157B near the 30 Doradus star formation
region in the LMC. The spin period (16 ms) and age ($\approx$ 5000
years) make it the fastest spinning young pulsar known. Scaling the
expected optical emission from the Crab pulsar, using the Pacini \&
Savati (1987) model, we would expect a pulsed magnitude in the range
$m_B \approx $ 19-21. Using archival CCD data, both Mignani et
al. (2000) and Gouiffes \& $\rm\ddot{O}$gelman (2001) show no obvious
counterpart candidate down to 23rd magnitude.

\subsection{TRIFFID Observations}

We observed the region around the ROSAT position in February 2000
using the ESO 3.6m telescope and the TRIFFID 2-d photometer. This
instrument consists of a relatively low quantum efficiency MAMA
detector, observing in B and V, in combination with a high sensitivity
(system QE 25 \%) group of avalanche photodiodes (APDs). The primary
APD was positioned towards the centre of the X-ray error circle on
candidate star 16 on Figure 2. Conditions were not photometric and the
seeing varied between 1.1'' and 1.6''. The region was observed on two
nights for a total of 5 hours. No pulsed signal was observed from the
APD at a limiting magnitude of $m_R \approx $ 24. A timing signal was
then searched for in the MAMA data, at the locations of all stars
marked within the 7''-radius circle in Figure 2; again no signal was
seen. A final search was then performed over the full area of the
7''-radius error circle using a set of multiplexed seeing width
apertures; again no signal at the pulsar frequency was observed. Our
limiting magnitude is estimated as $ m_B$ = 23.6 at the 3 $\sigma$
level. This upper limit implies a pulsed optical / X-ray ratio of
about $10^{-3}$.

\begin{figure}
\centerline{\psfig{file=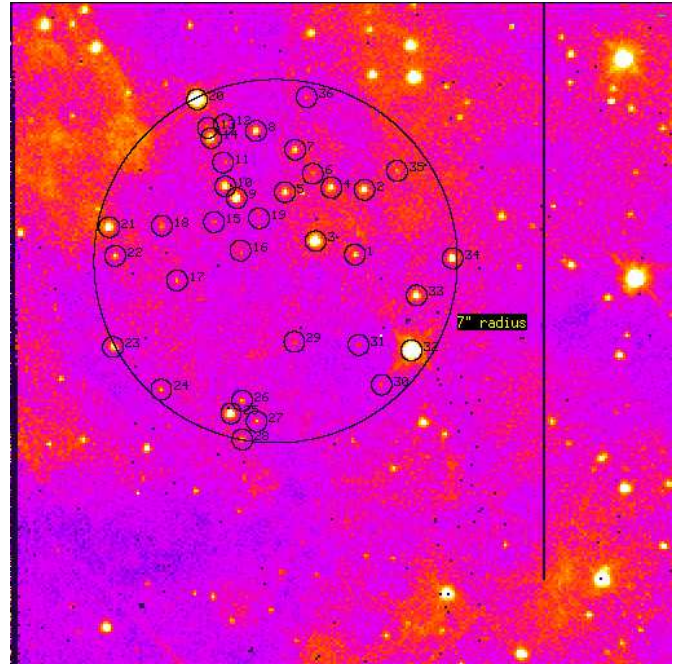,width=8.8cm,clip=} }
\caption{Section of the HST/WFPC2 F606W image (see Section 4.2) of N157B. 
This shows the circular region, of 7'' radius and centered on the
ROSAT position for PSR J0537-6910, which was searched by the TRIFFID
photometer.
\label{image}}
\end{figure}

\begin{figure}
\centerline{\psfig{file=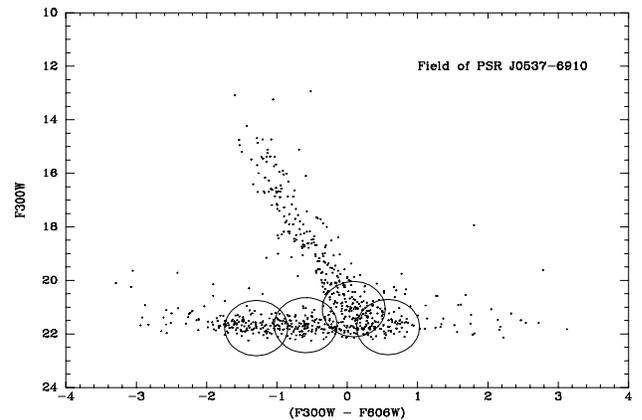,width=8.8cm,clip=} }
\caption{The HST/WFPC2 colour magnitude diagram (in F300W,
F300W-F606W) for stars in the N157B field. Circles mark the stars
within the Chandra error circle for PSR J0537-6910 (see Figure 4
below). The bluewards tail at F300W$\approx$22 is an artefact of the
different depths in the two filters.
\label{image}}
\end{figure}

\subsection{The Chandra-HST registration of the N157B field}
 
We obtained from the HST archive all WFPC2 images which overlapped the
field of N157B. The only moderately deep exposures where in the F606W
filter (``wide V-band''). Again, image processing, astrometry, and
photometry were performed with STSDAS, IRAF, and DAOPHOT-II. We
searched for possible pulsar optical counterpart candidates amongst
the stars detected within the ROSAT X-ray error circle for PSR
J0537-6910, centred at RA 05:37:47.1 Dec -69:10:23.0, by mapping this
coordinate into the astrometric solution in the WFPC2 file
headers. Figure 3 shows the colour-magnitude diagram for the stars in
the field. A similar analysis, using the RA \& Dec derived by Wang et
al. (2001) for the pulsar in their Chandra observations, yielded a
similar optical (mapped) position, albeit with a somewhat smaller
error circle.  However a comparison with the same analysis by Wang et
al. (2001), shown in their Fig. 2, shows a striking positional
difference of approx. 7 - 8 arcsec. This is very difficult to explain,
as the Chandra coordinate for the pulsar should be good to 1 arcsec,
based on agreement of this order with the 2 Wolf-Rayet stars in the
X-ray field; and the average absolute pointing error for the HST
pipeline astrometric solution (based on the Guide Star Catalog) is 0.8
- 1.5 arcsec (Biretta et al. 2000). Furthermore, our astrometry on the
HST image (shown in Figure 4) agrees very well with our independent
astrometry on NTT-SUSI imagery (Gouiffes \& $\rm\ddot{O}$gelman, 2001). One must
conclude that some error was made in the Chandra-HST registration of
the N157B field by Wang et al. (2001).

Indeed, if the X-ray contours in Fig. 2 of Wang et al. (2001) were
shifted to agree with our computed position, then these X-ray contours
would match the underlying nebulosity structures much better. This
would also have the effect of weakening their argument regarding the
extended X-ray emission extending ``beyond [filament] F4'' - the extent
would drop back by about 2 contour levels - although it is not
entirely incorrect. More seriously, their statement that ``the image
shows no evidence for the optical counterpart of the pulsar'' surely
becomes invalid, as they apparently searched the wrong area of the HST
image.

Although we can better address the latter question, having searched
the correct area of the HST image, we cannot report a convincing
optical counterpart either, because the HST exposures were not deep
enough - especially in the F300W filter (approx. U-band), which
prevented us from obtaining colour indices (~U-V) of the fainter F606W
sources. This illustrates the need for deeper, high-resolution imaging
and time-resolved imaging. The TRIFFID photometer would be capable of
reaching 26th magnitude with 10 hours of observation under reasonable
conditions.

\begin{figure}
\centerline{\psfig{file=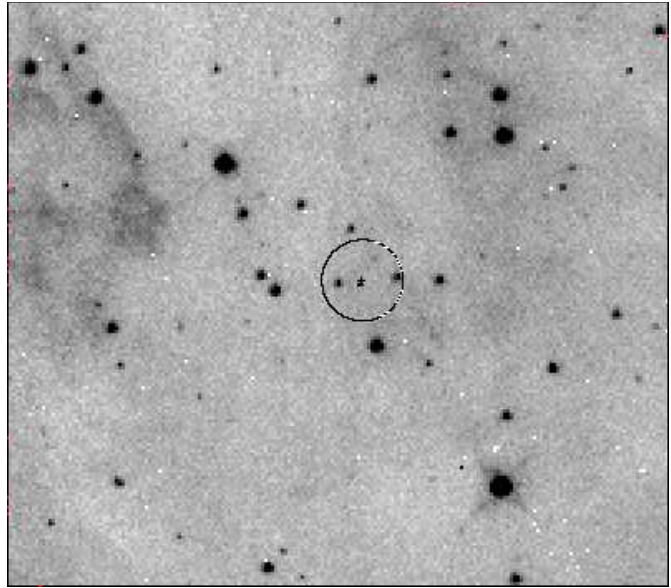,width=8.8cm,clip=} }
\caption{Aligned HST/WFPC2 field and Chandra X-ray error circle for PSR J0537-6910, from this work.
\label{image}}
\end{figure}

\begin{acknowledgements}
The authors gratefully acknowledge financial support from Enterprise
Ireland under the Basic Research Programme. RFB is also grateful for
financial support from the Improving Human Potential programme of the
European Commission (contract HPFM-CT-2000-00652). This publication is
partly based upon Hubble Space Telescope data obtained from the ST-ECF
archive, ESO, Garching, Germany. It also makes use of data products
from the Two Micron All Sky Survey, which is a joint project of the
University of Massachusetts and the Infrared Processing and Analysis
Center/California Institute of Technology, funded by the National
Aeronautics and Space Administration and the National Science
Foundation. Finally, we also wish to thank Richard Strom
for useful suggestions and comments while we prepared this paper.
\end{acknowledgements}
   

%
\clearpage

\end{document}